\documentclass[useAMS,usenatbib,fleqn]{mnras}
\usepackage{newtxtext,newtxmath}
\usepackage{algorithm,algorithmic}
\usepackage[T1]{fontenc}
\usepackage{ae,aecompl}

\usepackage[dvips]{graphicx}
\usepackage{epsfig,amssymb,amsmath,color}
\newcommand{\beq}{\begin{equation}}
\newcommand{\eeq}{\end{equation}}
\newcommand{\beqn}{\begin{eqnarray}}
\newcommand{\eeqn}{\end{eqnarray}}
\DeclareMathOperator*{\argmin}{arg\,min}
\def\bmath#1{\mbox{\boldmath$#1$}}

\long\def\symbolfootnote[#1]#2{\begingroup%
\def\thefootnote{\fnsymbol{footnote}}\footnote[#1]{#2}\endgroup}

\usepackage{pstricks,pst-text}

\title[Statistical Performance of Calibration]{Statistical Performance of Radio Interferometric Calibration}
\author[Yatawatta]{Sarod Yatawatta\\
ASTRON, Postbus 2, 7990 AA Dwingeloo, the Netherlands}

\begin{document}
\date{\today}
\pagerange{\pageref{firstpage}--\pageref{lastpage}} \pubyear{2018}
\maketitle
\label{firstpage}
\begin{abstract}
Calibration is an essential step in radio interferometric data processing that corrects the data for systematic errors and in addition, subtracts bright foreground interference to reveal weak signals hidden in the residual. These weak and unknown signals are much sought after to reach many science goals but the effect of calibration on such signals is an ever present concern. The main reason for this is the incompleteness of the model used in calibration. Distributed calibration based on consensus optimization has been shown to mitigate the effect due to model incompleteness by calibrating data covering a wide bandwidth in a computationally efficient manner. In this paper, we study the {\em statistical} performance of direction dependent distributed calibration, i.e., the distortion caused by calibration on the residual statistics. In order to study this, we consider the mapping between the input uncalibrated data and the output residual data. We derive an analytical relationship for the {\em influence} of the input on the residual and use this to find the relationship between the input and output probability density functions. The eigenspectrum of the Jacobian of this mapping is a direct indicator of the statistical performance of calibration. The analysis developed in this paper can also be applied to other data processing steps in radio interferometry such as imaging and foreground subtraction as well as to many other machine learning problems.
\end{abstract}
\section{Introduction}
Many challenging science cases in modern radio astronomy are after weak signals that are hidden under noise and bright foregrounds (Exoplanets \cite{Zarka2010}; Epoch of Reionization \cite{SZ2013}; Cosmic Dawn \cite{Singh2015} etc.). In order to reach these goals, the data need to be corrected for systematic errors and moreover, bright foreground interference need to be subtracted. We call both these operations as {\em calibration} and accurate and statistically unbiased calibration is essential for almost all radio interferometric observations. Consensus optimization \citep{boyd2011} has proven to be a computationally efficient  manner of calibrating radio interferometric data \citep{DCAL,EUSIPCO2016,Brossard2018,CAMSAP2017,DMUX,OLLIER2018} and results based on real observations \citep{Patil2017,Bharat2018} have already confirmed the enhanced accuracy. 

The ground truth sky model is hardly used in calibration either because of the computational complexity in evaluating such a model or simply because we never have the full information about the sky. Therefore, calibration is accurate only up to a certain extent. Because of this reason, performance measures of calibration are essential in order to build confidence in the end result which delivers the science. There are many ways to analyze the performance of calibration. Using signal estimation theory, Cramer-Rao lower bounds (CRLB) \citep{Zmu,Jeffs06,Wijn,Kazemi12} have been used to study the asymptotic variance of estimation error of calibration parameters. However, calibration estimates the systematic errors in the data and not the weak signals of scientific interest themselves. The weak signals are hidden in the residual and translating the variance bounds to the error in the residual is cumbersome. In order to overcome this, calibration can be considered as a nonlinear regression and Jacobian leverage \citep{cook1982residuals,Laurent92,Laurent93} is used in \citep{SS9,Patil2016} to directly get limits on the variance of the residuals. The performance of calibration can be measured during image formation as well. The increased variance of the residuals due to imperfect calibration is used by \cite{Bonn2018} to weight the data during image formation. Coherent errors in calibration are studied as image artifacts by \cite{grobler2014}. Power spectra of residual data are used to study weak signals that can only be detected by measuring statistical anomalies. The effect of imperfect calibration on power spectra are studied in e.g., \citep{Barry2016,EW2017,Millad2018,Joseph2018}, with various approximations to calibration thus only serving as guidelines.

In this paper, we analyze the performance of direction dependent distributed calibration using consensus optimization \citep{DCAL}. Our analysis (which is an extension of our previous work  \citep{SAM2018}) is general and  can easily be adopted for specific calibration scenarios such as traditional calibration (without using consensus) or calibration using a scalar data model (single polarization) or direction independent calibration. Furthermore, the same method of analysis can be used to study the performance of other data processing steps in radio interferometry such as imaging  \citep{Onose,Meil2016,Degu2016,Onose2017}  or foreground removal \citep{Chapman2013,CHIPS,Mertens2018}. In addition, the same analysis can be applied to other machine learning problems as well, for instance to study over fitting \citep{Tetko1995,srivastava14a}. As reported initially in \citep{SAM2018}, we consider the mapping between the input uncalibrated data and the output residual data, where the residual is obtained after calibration and removal of the bright foreground signals. We also consider correction of the data using the calibration solutions but only as a minor deviation from our analysis. We have derived an analytic relationship for the {\em influence} of the input data on the residual. Statistically speaking, the influence gives us a measure of robustness of calibration to outliers in the data \citep{Hempel86}. Lower influence indicates that the residual is less affected by outliers, in particular, unmodeled signals in the data such as diffuse foregrounds. Therefore, we can use influence as a direct measure of model errors in calibration.

The statistical performance of calibration is a measure of how much the input and residual noise statistics differ. In other words, it is a measure of the distortion caused by calibration on the residual noise statistics. We use influence to relate the probability density functions (PDFs) of the input raw data and the output residual. Note that we primarily use our analysis to study the performance of radio interferometric calibration assuming that calibration can be performed to a satisfactory level. Our previous work have already studied the selection of regularization and consensus polynomials for best calibration performance \citep{EUSIPCO2016,CAMSAP2017,DMUX} to reach this level. In this paper, we show that it is possible to measure the statistical performance of radio interferometric calibration (over fitting, increased variance etc.) in the presence of model errors by using the eigenspectrum of the Jacobian of the mapping between the input and the output. This enables us to process large volumes of radio interferometric data without manual checking and attach a confidence level to the output of calibration. When manual checking is feasible, it is also possible to create {\em influence maps} of the data. Unlike the effects caused by noise (which generally average to zero), the influence due to model error creates a systematic effect which can be observed by making images of the averaged influence.

The rest of the paper is organized is as follows. In section \ref{sec:model}, we summarize the performance analysis we have already developed and extend it to direction dependent calibration. In section \ref{sec:case}, we present several case studies on the application of our analysis to study the performance of well known regression problems. We use a simulated radio interferometric observation in section \ref{sec:simul} to test our performance analysis before drawing our conclusions in section \ref{sec:conc}.

{\em Notation}: Lower case bold letters refer to column vectors (e.g. ${\bf y}$). Upper case bold letters refer to matrices (e.g. ${\bf C}$). Unless otherwise stated, all parameters are complex numbers. The set of complex numbers is given as ${\mathbb C}$ and the set of real numbers as  ${\mathbb R}$. The matrix inverse, pseudo-inverse, transpose, Hermitian transpose, and conjugation are referred to as $(\cdot)^{-1}$, $(\cdot)^{\dagger}$, $(\cdot)^{T}$, $(\cdot)^{H}$, $(\cdot)^{\star}$, respectively. The matrix Kronecker product is given by $\otimes$. The vectorized representation of a matrix is given by $\mathrm{vec}(\cdot)$. The $i$-th element of a vector ${\bf y}$ is given by $y_i$. The identity matrix of size $N$ is given by ${\bf I}_N$.  Estimated parameters are denoted by a hat, $\widehat{(\cdot)}$. All logarithms are to the base $e$, unless stated otherwise. The Frobenius norm is given by $\|\cdot \|$. For a scalar function $f({\bmath \theta})$ of parameter vector ${\bmath \theta}$ (length $M$), $\frac{\partial f({\bmath \theta})}{\partial {\bmath \theta}}$ gives a column vector of derivatives (length $M$)  while $\frac{\partial f({\bmath \theta})}{\partial {\bmath \theta}^T}$ gives a row vector of derivatives (length $M$).

\section{Performance of Radio Interferometric Calibration}\label{sec:model}
We give a brief overview of radio interferometric calibration using consensus optimization \citep{DCAL} and summarize the performance bounds derived in \citep{SAM2018}. Consider  an array of $N$ receiving elements with dual polarized feeds and the correlation of signals at the $p$-th receiver and the $q$-th receiver, at frequency $f$, with proper signal delay, gives \citep{HBS} 
\beq \label{vispq}
{\bf { V}}_{pqf}= {\bf { J}}_{pf} {\bf { C}}_{pqf} {\bf { J}}_{qf}^{H} + {\bf { N}}_{pqf}.
\eeq
In (\ref{vispq}), ${\bf { V}}_{pqf}$ ($\in {\mathbb C}^{2\times 2}$) is the observed data (also called as the {\em visibilities}), and ${\bf { J}}_{pf}$,${\bf { J}}_{qf}$ ($\in {\mathbb C}^{2\times 2}$)  are the Jones matrices describing systematic errors at frequency $f$, at stations $p$ and $q$, respectively. The Jones matrices represent the effects of the propagation medium, the beam shape and the receiver. The noise matrix is given by ${\bf { N}}_{pqf}$ ($\in {\mathbb C}^{2\times 2}$). The intrinsic signal on baseline $pq$ is given by the coherency matrix ${\bf { C}}_{pqf}$ ($\in {\mathbb C}^{2\times 2}$) that can be pre-calculated by using the sky model \citep{TMS}.

In traditional calibration, the cost function that is minimized (under Gaussian noise) is given as
\beq \label{cost}
g_{f}({\bf J}_f)= \sum_{p,q}\| {\bf V}_{pqf} - {\bf A}_p{\bf J}_f {\bf C}_{pqf} ({\bf A}_q{\bf J}_f)^H \|^2
\eeq
where the systematic errors for all $N$ stations are grouped into ${\bf J}_f\in \mathbb{C}^{2N\times 2}$,
\beq \label{Jblock}
{\bf J}_f\buildrel\triangle\over=[{\bf J}_{1f}^T,{\bf J}_{2f}^T,\ldots,{\bf J}_{Nf}^T]^T.
\eeq
Using the canonical selection matrix ${\bf A}_p$ ($\in \mathbb{R}^{2\times 2N}$), where only the $p$-th block is ${\bf I}_2 \in \mathbb{R}^{2\times 2}$,
\beq \label{Ap}
{\bf A}_p \buildrel\triangle\over=[{\bf 0},{\bf 0},\ldots,{\bf I}_2,\ldots,{\bf 0}],
\eeq
we can select from ${\bf J}_f$ the systematic errors for the station $p$ as ${\bf A}_p{\bf J}_f$.
Note that in (\ref{cost}), the summation is taken over all the baselines $pq$ that are being calibrated, within a small bandwidth and time interval within which the systematic errors are assumed to be fixed.

As we have shown before, traditional calibration by solving (\ref{cost}) gives inferior results \citep{DCAL}. In contrast, distributed calibration calibrates data at all frequencies together using consensus optimization and is formulated as follows. First, we create the augmented Lagrangian as
\beq \label{aug}
L_f({\bf J}_f,{\bf Z},{\bf Y}_f)=g_{f}({\bf J}_f) + \|{\bf Y}_f^H({\bf J}_f-{\bf B}_f {\bf Z})\| + \frac{\rho}{2} \|{\bf J}_f-{\bf B}_f {\bf Z}\|^2
\eeq
where the subscript $(\cdot)_f$ denotes data (and parameters) at frequency $f$. In (\ref{aug}), $g_{f}({\bf J}_f)$ is the original cost function as in (\ref{cost}). The Lagrange multiplier is given by ${\bf Y}_f$ ($\in \mathbb{C}^{2N\times 2}$). The global variable ${\bf Z}$ ($\in \mathbb{C}^{2FN\times 2}$) is shared by data at all $P$ frequencies. Consensus polynomial basis (with $F$ terms) is represented by the matrix ${\bf B}_f = {\bf b}_f^T\otimes {\bf I}_{2N}$ ($\in \mathbb{R}^{2N\times 2FN}$) with ${\bf b}_f$ ($\in \mathbb{R}^{F\times 1}$) representing the basis functions evaluated at frequency $f$. The regularization parameter is given by $\rho$ ($\in \mathbb{R}^{+}$).

The alternating direction method of multipliers (ADMM) iterations $n=1,2,\ldots$ for solving (\ref{aug}) are given as
\beqn \label{step1}
({\bf J}_f)^{n+1}= \underset{{\bf J}}{\argmin}\ \ L_f({\bf J},({\bf Z})^n,({\bf Y}_f)^n)\\ \label{step2}
({\bf Z})^{n+1}= \underset{{\bf Z}}{\argmin}\ \ \sum_f L_f(({\bf J}_f)^{n+1},{\bf Z},({\bf Y}_f)^n)\\ \label{step3}
({\bf Y}_f)^{n+1}=({\bf Y}_f)^n + \rho\left( ({\bf J}_f)^{n+1}-{\bf B}_f ({\bf Z})^{n+1} \right)
\eeqn
where we use the superscript $(\cdot)^n$ to denote the $n$-th iteration where (\ref{step1}) to (\ref{step3}) are executed in order. The steps (\ref{step1}) and (\ref{step3}) are done for each $f$ in parallel using a network of computers, thus distributing the compute load. The update of the global variable in (\ref{step2}) is a simple linear operation and it can be done in closed form at the fusion center. 

Once we have the calibration solutions ${\bf J}_f$, we perform two major operations to the data.
First, we can calculate the residual ${\bf R}_{pqf}$ ($\in \mathbb{C}^{2\times 2}$) by subtracting the calibrated model from the input data as
\beq \label{residual}
{\bf R}_{pqf}={\bf V}_{pqf} - {\bf A}_p {\bf J}_f {\bf C}_{pqf} {\bf J}_f^H {\bf A}_q^T.
\eeq
On the other hand, we can also correct the input data to get
\beq \label{corr}
{\bf Q}_{pqf}= ({\bf A}_p {\bf J}_f)^{-1} {\bf V}_{pqf}  ({\bf A}_q{\bf J}_f)^{-H}
\eeq
where we call ${\bf Q}_{pqf}$ ($\in \mathbb{C}^{2\times 2}$) as the corrected data. We can also combine (\ref{corr}) and (\ref{residual}) as a cascade of operations and study the overall performance but we leave that out in our analysis for simplicity. The residual (\ref{residual}) can also be obtained after calibration along multiple directions as we describe later.

The main idea behind our performance analysis is as follows. We consider a small change in the input and find the corresponding change in the residual or in the corrected data. In statistical terms, we measure the {\em influence} \citep{cook1982residuals} of each input data point on the output residual or the corrected data. At convergence (gradient of the cost function with respect to the parameters is zero), the relative change between the input and the output (either the residual or the corrected data) should be as small as possible. The lower the influence, the more robust calibration is to outliers in the data \citep{Hempel86}. Not only does the influence indicate the performance of the calibration algorithm used, but it also gives a measure of how well we have formulated our calibration problem, i.e., the sky model we have used and the noise model we have assumed. 

Consider one input data point out of many that form the full observation. The visibility matrix ${\bf V}_{pqf}$  in (\ref{vispq}) is composed of $4$ complex values or $8$ real values for each $p$,$q$ and $f$. We consider one such data point as $x_{p^\prime q^\prime r} \in \mathbb{R}$ and we have implicitly assumed the $f$ dependence for simplicity of notation. This data point belongs to $p=p^\prime,q=q^\prime$ baseline and $r\in[1,8]$ (and we consider $f$ to be the same). We select the value of $r$ to represent one real or one imaginary value of ${\bf V}_{p^\prime q^\prime f}$. Note that each complex number is considered as two data points. For instance, if $r=1$, we represent the real part of ${\bf V}_{p^\prime q^\prime f}(1,1)$. If $r=2$, the imaginary part of ${\bf V}_{p^\prime q^\prime f}(1,1)$ is selected, and so on.

Using techniques developed in \citep{Mdiff,Samuel,Gould2016}, we have derived the expressions for the influence of $x_{p^\prime q^\prime r}$ on the residual and the corrected data in \citep{SAM2018}. We have the influence on the residual as
\beqn \label{Rderiv}
 \mathrm{vec}\left( \frac{\partial{\bf R}_{pqf}} {\partial x_{p^\prime q^\prime r}} \right)=&& \mathrm{vec}\left( \frac{\partial{\bf V}_{pqf}} {\partial x_{p^\prime q^\prime r}} \right)\\\nonumber
&&-\left({\bf C}_{pqf}{\bf J}_f^H {\bf A}_q^T\right)^T \otimes {\bf A}_p \mathrm{vec}\left( \frac{\partial{\bf J}_{f}} {\partial x_{p^\prime q^\prime r} } \right).
\eeqn
Note that $\mathrm{vec}\left( \frac{\partial{\bf V}_{pqf}} {\partial x_{p^\prime q^\prime r}} \right)$ in (\ref{Rderiv}) is zero except when $p=p^\prime,q=q^\prime$ (and $f$ is the same).
Similarly, the influence of $x_{p^\prime q^\prime r}$ on the corrected data is 
\beq \label{Qderiv}
\mathrm{vec}\left( \frac{\partial{\bf Q}_{pqf}} {\partial x_{p^\prime q^\prime r}} \right)= - {\bf Q}_{pqf}^T \otimes \left(({\bf A}_p {\bf J}_f)^{-1} {\bf A}_p\right) \mathrm{vec}\left( \frac{\partial{\bf J}_{f}} {\partial x_{p^\prime q^\prime r} } \right)
\eeq
and the proof is given in appendix \ref{app:Corr}.

The closed form expressions (\ref{Rderiv}) and (\ref{Qderiv}) are dependent on
\beqn \label{Jderiv}
\lefteqn{\mathrm{vec}\left(\frac{\partial {\bf J}_f}{\partial x_{p^\prime q^\prime r}}\right)}\\\nonumber
&&=\left( \mathcal{D}_{\bf J}{\rm grad}(g_{f}({\bf J}_f))\right. \\\nonumber
&&\left. + \frac{\rho}{2}{\bf I}_2\otimes\left({\bf F}^H{\bf F}\left({\bf I}_{2N}+\left({\bf I}_{2N}-{\bf F}^H{\bf F}\right)^{-1}{\bf F}^H{\bf F}\right)\right)\right)^{-1}\\\nonumber
&&\times \left({\bf A}_{q^\prime}{\bf J}_f {\bf C}_{p^\prime q^\prime f}^H\right)^T \otimes {\bf A}_{p^\prime}^T \mathrm{vec}\left(\frac{\partial {\bf V}_{p^\prime q^\prime f}}{\partial x_{p^\prime q^\prime r}}\right)
\eeqn
where ${\bf F}$  ($\in \mathbb{C}^{2N\times 2N}$) is a matrix which is exclusively formed by the consensus polynomial used in (\ref{aug}) and is independent of $\rho$ \citep{SAM2018}. 
The Hessian of the cost function $g_{f}({\bf J}_f)$ is given as
\beqn \label{DJ}
\lefteqn{\mathcal{D}_{\bf J}{\rm grad}(g_{f}({\bf J}_f))=}\\\nonumber
&&\sum_{p,q} \left( -({\bf C}_{pqf}^{H})^{T}\otimes {\bf A}_p^T {\bf R}_{pqf}{\bf A}_q
-{\bf C}_{pqf}^{T}\otimes {\bf A}_q^T {\bf R}_{pqf}^H{\bf A}_p  \right. \\\nonumber
&+&\left.({\bf C}_{pqf}{\bf J}_f^H{\bf A}_q^T{\bf A}_q{\bf J}_f{\bf C}_{pqf}^H)^T\otimes {\bf A}_p^T{\bf A}_p \right. \\\nonumber
&+&\left.({\bf C}_{pqf}^H{\bf J}_f^H{\bf A}_p^T{\bf A}_p{\bf J}_f{\bf C}_{pqf})^T\otimes {\bf A}_q^T{\bf A}_q \right) \\\nonumber
\eeqn
where we express it as a function of ${\bf R}_{pqf}$ and not as a function of input ${\bf V}_{pqf}$. This makes it easier to analyze calibration along multiple directions as we show later.

We can make several key conclusions using (\ref{Jderiv}) and (\ref{DJ}):
\begin{itemize}
\item By making $\rho=0$ in (\ref{Jderiv}), we get the performance of traditional calibration without consensus (i.e., data at each frequency are calibrated independently), but of course this assumes that the solutions are the same for both cases. We also see that by increasing $\rho$, the relative change of the solutions due to the input goes down.
\item The value of ${\bf F}$ in (\ref{Jderiv}) depends only on the consensus polynomials ${\bf B}_f$  chosen in (\ref{aug}). Therefore, we can use (\ref{Jderiv}) to evaluate the performance of various polynomial basis functions used to construct ${\bf B}_f$ (in addition to the method in \citep{DMUX}).
\item Consider a (hypothetical) perfect calibration, where ${\bf R}_{pqf}={\bf 0}$ and ${\bf J}_f$ is known. Even in this case, for a source with non zero ${\bf C}_{pqf}$, we see that $\mathcal{D}_{\bf J}{\rm grad}(g_{f}({\bf J}_f))$ in (\ref{DJ}) is not zero. The value of (\ref{Jderiv}) for this hypothetical scenario is the theoretical bound in $\frac{\partial {\bf J}_f}{\partial x_{p^\prime q^\prime r}}$ that we can achieve with perfect calibration.
\end{itemize}

The residual of a typical radio interferometric observation is obtained after calibration along many directions (or clusters of sources \citep{SS4}) in the sky, not just one direction. Therefore, we extend the result (\ref{Rderiv}) to multi-directional calibration as follows. Consider calibration along $K$ directions, where the direction indices $\{1,2,\ldots,K\}$ are represented as the set $\mathcal{C}$. We find the residual by subtracting a set of directions whose indices are in the set $\mathcal{S}$. Note that $\mathcal{S}\subseteq \mathcal{C}$ and in most cases $\mathcal{S}=\mathcal{C}$. However, there are exceptions to this. For instance, we might be interested in preserving the signal along one direction in the sky (where the scientific interest lies) while removing sources in all other directions. In this case, we calibrate along $K$ directions but find the residual by only subtracting signals from $K-1$ directions while preserving the signal along the interesting direction. Another example where $\mathcal{S}\ne \mathcal{C}$ is when we only have an approximate model for certain signals in the sky. For instance, we might have an approximate model for the diffuse, large scale structure in the sky and we can use this approximate model in calibration but we do not subtract this from the data to find the residual. 

We use superscripts $(\cdot)^k$ and $(\cdot)^l$ to denote $k$-th and $l$-th directions, respectively. At convergence, the effective data seen by the $k$-th direction ($k\in \mathcal{C}$) is \citep{Kaz2}
\beq \label{Vem}
{\bf { V}}^{k}_{pqf}= {\bf { V}}_{pqf} - \sum_{l\in \mathcal{C},l\ne k}{\bf { J}}^l_{pf} {\bf { C}}^l_{pqf} ({\bf { J}}^l_{qf})^{H}. 
\eeq
The solution can be found by minimizing the cost function (\ref{cost}) per each direction by using the expectation maximization (EM) algorithm \cite{Kaz2}. At convergence, any other method of calibration will also lead to the same solution and therefore the following derivations based on the EM algorithm hold for any other method of calibration. The residual seen by the $k$-th direction at convergence is
\beq \label{Rem}
{\bf { R}}^{k}_{pqf}={\bf { V}}^{k}_{pqf} - {\bf { J}}^k_{pf} {\bf { C}}^k_{pqf} ({\bf { J}}^k_{qf})^{H}
\eeq
and substituting ${\bf { V}}^{k}_{pqf}$ in (\ref{Vem}) to (\ref{Rem}) we see that ${\bf { R}}^{k}_{pqf}$ is the same for all $k \in \mathcal{C}$. Hence we have a common residual ${\bf { R}}_{pqf} = {\bf { R}}^{k}_{pqf}$ for all $k$. Moreover, in evaluating (\ref{DJ}) for the $k$-th direction, only the coherencies ${\bf { C}}^{k}_{pqf}$ and the solutions ${\bf { J}}^k_{f}$ depend on $k$.

Now consider the case where $\mathcal{S}\ne \mathcal{C}$, i.e., the residual produced as the output is not equal to the residual used by the calibration algorithm. In that case, the residual produced as the output is
\beq \label{Reff}
\widetilde{\bf { R}}_{pqf}={\bf { V}}_{pqf} - \sum_{l\in \mathcal{S}}{\bf { J}}^l_{pf} {\bf { C}}^l_{pqf} ({\bf { J}}^l_{qf})^{H}
\eeq
and using the same indices $p^\prime, q^\prime, r$ as in (\ref{Rderiv}) and using the linearity of differentiation, we get
\beqn \label{Reffderiv}
 \mathrm{vec}\left( \frac{\partial{\widetilde{\bf R}}_{pqf}} {\partial x_{p^\prime q^\prime r}} \right)&& =\mathrm{vec}\left( \frac{\partial{\bf V}_{pqf}} {\partial x_{p^\prime q^\prime r}} \right)\\\nonumber
&&-\sum_{k \in \mathcal{S}} \left({\bf C}^k_{pqf}({\bf J}^k_f)^H {\bf A}_q^T\right)^T \otimes {\bf A}_p \mathrm{vec}\left( \frac{\partial{\bf J}^k_{f}} {\partial x_{p^\prime q^\prime r} } \right).
\eeqn
Note that the residual used to evaluate $\left( \frac{\partial{\bf J}^k_{f}} {\partial x_{p^\prime q^\prime r} } \right)$ in (\ref{Reffderiv}) is given by (\ref{Rem}) and is not equal to $\widetilde{\bf { R}}_{pqf}$ when $\mathcal{S}\ne \mathcal{C}$. Otherwise, the expressions needed to evaluate (\ref{Reffderiv}) are similar to the single direction case as given in (\ref{Jderiv}) and (\ref{DJ}).

The expressions presented thus far are in a space of complex matrices, and we convert them into expressions in a space of real vectors \citep{SAM2018} to study the statistical relationships between the input data and the output residual.
We consider a model 
\beq \label{obs}
{\bf x}={\bf s}({\bmath \theta}) + {\bf n}
\eeq
where ${\bmath \theta}$ ($\in \mathbb{R}^{M\times 1}$) is the real parameter vector that is estimated by calibration. The elements of ${\bmath \theta}$ are the elements of ${\bf { J}}^k_{pf}$-s, with real and imaginary parts considered separately. The input data ${\bf x}$  ($\in \mathbb{R}^{D\times 1}$) is composed of the values of ${\bf { V}}_{pqf}$ in (\ref{vispq}). The model ${\bf s}({\bmath \theta})$ is a mapping between $\mathbb{R}^{M\times 1}$ and $\mathbb{R}^{D\times 1}$.

 For a single time sample, $D=8N(N-1)/2$ because each (unique) cross correlation produces $8$ real data points. However, $D$ can be larger than this value if more than one time sample is calibrated to obtain a common solution. One element of ${\bf x}$ is  $x_{p^\prime q^\prime r}$ which was considered in (\ref{Rderiv}) and in (\ref{Qderiv}). To shorten our notation, we use $x_m$ to represent $x_{p^\prime q^\prime r}$ where $m$ is the $m$-th element of ${\bf x}$. Note that we always have a one-to-one (bijective) mapping between $p^\prime q^\prime r$ and $m$.

Calibration is essentially minimizing a cost function $f({\bf x},{\bmath \theta})$,
\beq \label{mltheta}
\widehat{\bmath \theta}=\underset{\bmath \theta}{\rm arg\ min} f({\bf x},{\bmath \theta}) 
\eeq
to get $\widehat{\bmath \theta}$ as our solution. Note that $f({\bf x},{\bmath \theta})$ need not have any direct relationship with the model ${\bf s}({\bmath \theta})$. For instance, we can have additional terms than just ${\bf s}({\bmath \theta})$ for regularization. Other noise models such as Student's t \citep{Kaz3} or spherically invariant random processes \citep{SIRP} will yield other more exotic cost functions.

The output ${\bf y}$ ($\in \mathbb{R}^{D\times 1}$) is the residual using the calibration solution $\widehat{\bmath \theta}$ 
\beq \label{residualv}
{\bf y}= {\bf x}-{\bf s}(\widehat{\bmath \theta})
\eeq
which can also be expressed as
\beq \label{Tx}
{\bf y}={\bf T}({\bf x})
\eeq
where ${\bf T}(\cdot)$ is a one-to-one mapping between the input ${\bf x}$ and the output ${\bf y}$.

Using (\ref{Tx}) we can relate the PDFs of the input and the output as \citep{fessler1998transformations}
\beq \label{xypdf}
p_X({\bf x}) = | \mathcal{J} | \ \ p_Y({\bf T}({\bf x}))  
\eeq
where $\mathcal{J}$ is the Jacobian of the mapping  ${\bf T}(\cdot)$, i.e.,
\beq \label{Jmapping}
 \mathcal{J} =\left[
\begin{array}{cccc}
\frac{\partial y_1}{\partial x_1} & \frac{\partial y_1}{\partial x_2} & \ldots & \frac{\partial y_1}{\partial x_D}\\
\frac{\partial y_2}{\partial x_1} & \frac{\partial y_2}{\partial x_2} & \ldots & \frac{\partial y_2}{\partial x_D}\\
\vdots & \vdots & \vdots & \vdots\\
\frac{\partial y_D}{\partial x_1} & \frac{\partial y_D}{\partial x_2} & \ldots & \frac{\partial y_D}{\partial x_D}
\end{array} \right].
\eeq

 As shown in \cite{SAM2018}, the determinant of the Jacobian of ${\bf T}(\cdot)$ can be expressed as \citep{Ipsen}
\beq \label{deteigs}
| \mathcal{J} | = \exp \left(\sum_{i=1}^{D} \log \left(1 + \lambda_i(\mathcal{ A})\right) \right).
\eeq
The value of $| \mathcal{J} |$ is entirely dependent on the values of $\lambda_i(\mathcal{A})$ which are the eigenvalues of $\mathcal{A}$ $\in \mathbb{R}^{D\times D}$,
\beq \label{AA}
 \mathcal{A}\buildrel \triangle\over=\frac{\partial {\bf s}({\bmath \theta})}{\partial {\bmath \theta}^T} \left( f_{\theta \theta} ({\bf x},{\bmath \theta}) \right)^{-1} [f_{X_1 \theta} ({\bf x},{\bmath \theta})\ldots f_{X_D \theta} ({\bf x},{\bmath \theta})] \arrowvert_{{\bmath \theta}=\widehat{\bmath \theta}},
\eeq
where
\beqn \label{fder}
f_{\theta \theta} ({\bf x},{\bmath \theta}) \buildrel \triangle \over= \frac{\partial^2 f({\bf x},{\bmath \theta})} {\partial {\bmath \theta} \partial{\bmath \theta}^T}   \in \mathbb{R}^{M\times M},\\
f_{X_m \theta} ({\bf x},{\bmath \theta}) \buildrel \triangle \over= \frac{\partial^2 f({\bf x},{\bmath \theta})}{\partial x_m \partial {\bmath \theta}}  \in \mathbb{R}^{M\times 1},
\eeqn
and $\frac{\partial {\bf s}({\bmath \theta})}{\partial {\bmath \theta}^T}  \in \mathbb{R}^{D\times M}$ is the derivative (or Jacobian) of the model ${\bf s}({\bmath \theta})$. 

Evaluating (\ref{Jmapping}) or (\ref{AA}) can be done in closed form using (\ref{Reffderiv}) for a general direction dependent calibration. There are several ways to minimize the computational cost for large $D$. First, we can select a subset of rows and columns from (\ref{Jmapping}) to consider a reduced one-to-one mapping (by selecting the same set of rows and columns, see appendix \ref{app:Subset}). Averaging is another possible option, where (\ref{Jmapping}) is averaged over a small time and frequency interval. Normally calibration is also performed over a small time and frequency interval and the same interval can be used to average entries in (\ref{Jmapping}). We can develop this averaging further by noting that $D$ data points create $D^2$ values in (\ref{Jmapping}) and if we can average this down to $D$ values, we can replace the residual ${\bf y}$ by the averaged influence values and feed it to imaging routines. An obvious way to do this is to average each row (or column) to a single value. Row averaging would give us the average influence of each output data point created by all input data points and column averaging would give us the converse. Considering each element of (\ref{Jmapping}) as a gradient, we see that averaging is a way to reduce variance due to noise and reveal systematic trends \citep{VarReduc}, especially due to outliers in the data. We call this averaging over rows and making images of the averaged influence as {\em influence mapping}.

The statistical distortion of the residual ${\bf y}$ with respect to the input ${\bf x}$ is given by (\ref{xypdf}). Finding this involves finding the eigenvalues of $\mathcal{A}$ in (\ref{AA}) (which is also dependent on ${\bf y}$) which is an expensive task for large $D$. Assume the ideal case where we directly observe the signal we want, i.e., ${\bf y}={\bf x}$. No calibration needs to be done and ${\bf T}(\cdot)$ is the identity mapping. In this case, $| \mathcal{J} |=1$ and $\lambda_i(\mathcal{A})=0$ (or $1+\lambda_i(\mathcal{A})=1$) for all $i$.
By considering the realistic situation to be a slight deviation from the ideal case (assuming we have already fine-tuned calibration for best practical performance), we can look at the  dominant eigenvalues (largest magnitude) of $\mathcal{A}$ and consider their deviation from the ideal case as a measure of the statistical distortion due to calibration. We can use iterative methods (such as the implicitly restarted Arnoldi method \citep{ARPACK}) to find the dominant eigenvalues and we do not have to construct the full matrix $\mathcal{A}$. The inversion of $\left( f_{\theta \theta} ({\bf x},{\bmath \theta}) \right)$ in (\ref{AA}) is not explicitly needed because we can also use iterative algorithms to solve a linear system of equations, for instance we can use Richardson iterations \citep{Richardson307} to replace the inversion of $\left( f_{\theta \theta} ({\bf x},{\bmath \theta}) \right)$.

We have derived expressions for the statistical performance of direction dependent radio interferometric calibration or for the distortion of output statistics due to calibration. The obvious question one can ask is how to invert this distortion to get back the original statistics. For instance, this distortion can be seen as mode mixing \citep{Hazelton} in residual power spectra. The inversion of (\ref{xypdf}) is complicated due to the large dimentionality of ${\bf x}$ and ${\bf y}$. In future work, we shall seek simplifications of this relationship, for instance by using copula theory \citep{EmpCopula}. 

We use the expressions given in this section to analyze the performance of radio interferometric calibration in section \ref{sec:simul}. Before doing that, in order to help build our understanding, we give two case studies of well known regression examples in section \ref{sec:case}.
\section{Case studies}\label{sec:case}
The objective of this section is twofold. First, we look at simple and well known regression examples to help build our understanding of the performance measures developed in section \ref{sec:model}. This enables us to compare the performance of radio interferometric calibration with, e.g., linear regression. Secondly, we show the utility of the  method of analysis presented in section \ref{sec:model} in analyzing the performance of other machine learning examples. In this respect, we consider the proposed method as an improvement to using the spectral norm of the Jacobian of the model ${\bf s}({\bmath \theta})$  as done by \cite{Sokolic2017}.
\subsection{Linear model}\label{sec:example}
We consider a simple example that is well studied, i.e., a linear system 
\beq \label{ex_data}
{\bf x}={\bf A}{\bmath \theta} + {\bf n}
\eeq
where ${\bf A}\in \mathbb{R}^{D\times M}$ is a known matrix of rank $M$, ${\bf x}\in \mathbb{R}^{D\times 1}$ is the observed data, and ${\bf n} \in  \mathbb{R}^{D\times 1}$ is the noise. There are many applications of this in radio astronomy as well, too numerous to mention here individually. 
We have a linear model
\beq \label{ex_model}
{\bf s}({\bmath \theta})={\bf A}{\bmath \theta}
\eeq
where ${\bmath \theta}\in \mathbb{R}^{M\times 1}$ is the unknown parameter vector.
Consider using a least squares cost function with regularization $\beta \in \mathbb{R}^{+}$ (ridge regression)
\beq \label{ex_cost}
f({\bf x},{\bmath \theta})=({\bf x}-{\bf A}{\bmath \theta})^T({\bf x}-{\bf A}{\bmath \theta}) + \beta {\bmath \theta}^T {\bmath \theta}
\eeq
to find ${\bmath \theta}$. The gradient of the cost function is
\beq \label{ex_grad}
\frac{\partial f({\bf x},{\bmath \theta})}{\partial {\bmath \theta}}=2\left(-{\bf A}^T {\bf x}+({\bf A}^T{\bf A} + \beta {\bf I}_M){\bmath \theta}\right)
\eeq
and equating this to ${\bf 0}$ gives the solution
\beq \label{ex_sol}
\widehat{\bmath \theta}=({\bf A}^T{\bf A}+\beta {\bf I}_M)^{-1}{\bf A}^T {\bf x}.
\eeq
In order to study the performance using (\ref{AA}), we need 
\beqn \label{ex_comp0}
\frac{\partial s({\bmath \theta})}{\partial {\bmath \theta}^T}={\bf A},\\ \label{ex_comp1}
\frac{\partial^2 f({\bf x},{\bmath \theta})}{\partial {\bmath \theta} \partial {\bmath \theta}^T}=2({\bf A}^T {\bf A}+\beta{\bf I}_M),\\\label{ex_comp2}
\frac{\partial^2 f({\bf x},{\bmath \theta})}{\partial x_m \partial {\bmath \theta}} =-2{\bf A}^T {\bf e}_m
\eeqn
where ${\bf e}_m \in \mathbb{R}^{M\times 1}$ is the canonical vector with a $1$ at the $m$-th location and otherwise zero.
Substituting (\ref{ex_comp0}), (\ref{ex_comp1}), and (\ref{ex_comp2}) to (\ref{AA}) we get
\beq \label{Aex}
 \mathcal{A}=-{\bf A}({\bf A}^T{\bf A}+\beta{\bf I}_M)^{-1}{\bf A}^T
\eeq
for this example.
When $\beta=0$, (\ref{Aex})  is the well known (negative) hat matrix or the projection matrix \citep{cook1982residuals}, that has $M$ eigenvalues with value $-1$ and the rest of the eigenvalues are zero (assuming $D>M$). Therefore, $| \mathcal{J} |=0$ and (\ref{xypdf})  is undefined \citep{fessler1998transformations}.  Hence it is better to study the eigenvalues $\lambda(\mathcal{A})$ rather than the determinant in this example. Our interest here is to study the statistics of the residual ${\bf x}-{\bf A}\widehat{\bmath \theta}$ and try to see how different it is from the statistics of the noise ${\bf n}$. Ideally we want $| \mathcal{J} |=1$ so all eigenvalues of $\mathcal{A}$ should be zero and $1+\lambda(\mathcal{A})$ should be equal to $1$. We can do this by making $\beta=\infty$ but this gives a practically unusable result because the solution $\widehat{\bmath \theta}$ becomes zero.

\begin{figure}
\begin{minipage}{0.98\linewidth}
\centerline{\epsfig{figure=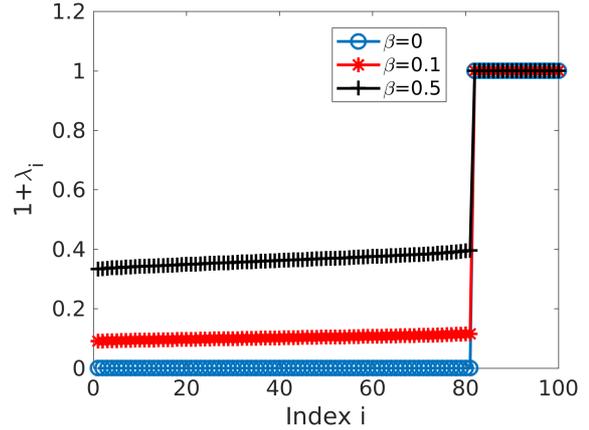,width=8.0cm}}
\end{minipage}
\caption{Ridge regression lowest $100$ values of $1+\lambda(\mathcal{A})$ for various values of regularization $\beta$.\label{ridge_eig}}
\end{figure}

  As a numerical example of ridge regression, we consider a source model construction using shapelet basis functions \citep{Shap}. We have an image of a radio source of size $128\times 128$ pixels, so $D=128\times 128=16384$. We use a rectangular shapelet basis with $M=9\times 9=81$ basis functions (normalized to have $\|{\bf A}\|=1$). We decompose the image into the $M$ basis to create a source model.  We have shown the eigenspectrum $1+\lambda(\mathcal{A})$ (for the $100$ dominant eigenvalues of $\mathcal{A}$) for various values of $\beta$ in Fig. \ref{ridge_eig}. 

As expected, when $\beta=0$, $\mathcal{A}$ has $M$ eigenvalues with value $-1$ and therefore $1+\lambda(\mathcal{A})=0$ as seen in Fig. \ref{ridge_eig}. The remaining $D-M$ values of $1+\lambda(\mathcal{A})$  are always $1$. Moreover, as the value of $\beta$ increases, we see that the values of $1+\lambda(\mathcal{A})$ are approaching $1$. For the extreme case of $\beta=\infty$, we get all $1+\lambda(\mathcal{A})=1$ but we will not get a usable model for the source as the solution $\widehat{\bmath \theta}$ becomes zero.

The selection of $\beta$ for best performance depends on the application, but we should note that if we are interested in studying the noise statistics by looking at the residual, any finite value of $\beta$ will result in a distortion. There is also the well known trade off between the bias and the variance of the residuals \citep{neco1992} that can be used to choose an appropriate value for $\beta$.

We extend the linear example (\ref{ex_data}) in a way analogous to multi frequency calibration as follows. Let the data taken at frequency $f_i$ be
\beq \label{exf_data}
{\bf x}_{f_i}={\bf A}_{f_i}{\bmath \theta} + {\bf n}
\eeq
where $i \in [1,P]$ and we have $P$ observations with models ${\bf A}_{f_i}$ that need not be the same.
The cost function is minimized over all data and is given by
\beq \label{exf_cost}
f({\bf x}_{f_1},\ldots,{\bf x}_{f_P},{\bmath \theta})=\sum_{i=1}^{P}({\bf x}_{f_i}-{\bf A}_{f_i}{\bmath \theta})^T({\bf x}_{f_i}-{\bf A}_{f_i}{\bmath \theta}) + \beta {\bmath \theta}^T {\bmath \theta}
\eeq

Now, we study the performance of data at $f=f_i$ out of the $P$ frequencies. The expressions analogous  to (\ref{ex_comp0}), (\ref{ex_comp1}) and (\ref{ex_comp2}) in this case are
\beqn \label{exf_comp0}
\frac{\partial s_{f}({\bmath \theta})}{\partial {\bmath \theta}^T}={\bf A}_{f},\\ \label{exf_comp1}
\frac{\partial^2 f({\bf x}_{f_1},\ldots,{\bf x}_{f_P},{\bmath \theta})}{\partial {\bmath \theta} \partial {\bmath \theta}^T}=2\left(\sum_{i=1}^{P}{\bf A}_{f_i}^T {\bf A}_{f_i}+\beta{\bf I}_M\right),\\\label{exf_comp2}
\frac{\partial^2 f({\bf x}_{f_1},\ldots,{\bf x}_{f_P},{\bmath \theta})}{\partial x_m \partial {\bmath \theta}} =-2{\bf A}_f^T {\bf e}_m
\eeqn
and we can study the performance of (\ref{exf_data}) by substituting (\ref{exf_comp0}), (\ref{exf_comp1}) and (\ref{exf_comp2}) to (\ref{AA}). Note that $x_m$ in (\ref{exf_comp2}) is assumed to be an element of ${\bf x}_{f}$.

\subsection{Quadratic model}\label{sec:example2}
We consider the simplest non-linear system, i.e., a quadratic model. The observed data ${\bf X}\in \mathbb{R}^{N\times M}$ is given as 
\beq \label{ex2_data}
{\bf X}={\bf A}{\bmath \theta}{\bmath \theta}^T + {\bf N}
\eeq
where ${\bf A}\in \mathbb{R}^{N\times M}$ is the model matrix, ${\bmath \theta} \in \mathbb{R}^{M\times 1}$ is the parameter vector and ${\bf N} \in \mathbb{R}^{N\times M}$ is the noise. The total number of data points observed is $D=N\times M$ and this can be written as a vector ${\bf x}=\mathrm{vec}({\bf X})$.

One major difference compared to the linear model in section \ref{sec:example} is that the solutions for ${\bmath \theta} \in \mathbb{R}^{M\times 1}$ need not be unique, i.e., if $\widehat{\bmath \theta}$ is a solution, then $-\widehat{\bmath \theta}$ is also a solution. 
The quadratic model is
\beq \label{ex2_model}
{\bf S}({\bmath \theta})={\bf A}{\bmath \theta}{\bmath \theta}^T,\  {\bf s}({\bmath \theta})= \mathrm{vec}({\bf S}({\bmath \theta}))
\eeq
with ${\bf s}({\bmath \theta}) \in \mathbb{R}^{D\times 1}$. We use the cost function
\beq \label{ex2_cost}
f({\bf X},{\bmath \theta})=\mathrm{trace}\left(({\bf X}-{\bf A}{\bmath \theta}{\bmath \theta}^T)^T({\bf X}-{\bf A}{\bmath \theta} {\bmath \theta}^T)\right) + \beta {\bmath \theta}^T {\bmath \theta}
\eeq
where $\beta\in \mathbb{R}^{+}$ is the regularization parameter.

After some matrix algebra, we get the gradient of the cost function as
\beq \label{ex2_grad}
\frac{\partial f({\bf X},{\bmath \theta})}{\partial {\bmath \theta}}=2\left(-{\bf X}^T{\bf A}-{\bf A}^T{\bf X}+{\bf A}^T{\bf A}{\bmath \theta}{\bmath \theta}^T +{\bmath \theta}{\bmath \theta}^T{\bf A}^T{\bf A} + \beta {\bf I}_M\right){\bmath \theta}
\eeq
and we do not have a closed form solution $\widehat{\bmath \theta}$ and need numerical optimization to minimize (\ref{ex2_cost}). Using the solution, we calculate the residual as
\beq\label{ex2_residual}
{\bf R}={\bf X}-{\bf A}\widehat{\bmath \theta}\widehat{\bmath \theta}^T.
\eeq

In order to study the performance, we need
\beq \label{ex2_sgrad}
\frac{\partial s({\bmath \theta})}{\partial {\bmath \theta}^T} = {\bmath \theta} \otimes {\bf A} + {\bf I}_M \otimes {\bf A}{\bmath \theta},
\eeq
and
\beqn \label{ex2_ftt}
\frac{\partial^2 f({\bf X},{\bmath \theta})}{\partial {\bmath \theta} \partial {\bmath \theta}^T} = 2\left(-{\bf R}^T {\bf A} - {\bf A}^T{\bf R}+\beta {\bf I}_M + {\bmath \theta}^T{\bmath \theta} \otimes {\bf A}^T{\bf A}\right.\\\nonumber
+\left.  {\bmath \theta}^T \otimes {\bf A}^T {\bf A}{\bmath \theta} +{\bmath \theta}^T {\bf A}^T {\bf A}{\bmath \theta} \otimes  {\bf I}_M + {\bmath \theta}^T {\bf A}^T {\bf A}\otimes {\bmath \theta} \right)
\eeqn
and 
\beq \label{ex2_ftx}
\frac{\partial^2 f({\bf X},{\bmath \theta})}{\partial x_m \partial {\bmath \theta}} =-2\left( \left( ({\bf A}{\bmath \theta})^T \otimes {\bf I}_M\right) {\bf K}_X + {\bmath \theta}^T \otimes {\bf A}^T \right){\bf e}_m
\eeq
for $x_m$ being one element from ${\bf x}$ ($m \in [1,D]$). Note that ${\bf K}_X$  in (\ref{ex2_ftx}) is the commutation matrix such that ${\bf K}_X \mathrm{vec}({\bf X})=\mathrm{vec}({\bf X}^T)$.

\begin{figure}
\begin{minipage}{0.98\linewidth}
\centerline{\epsfig{figure=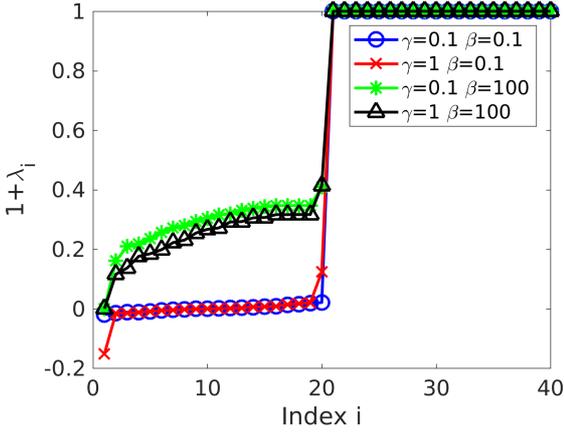,width=8.0cm}}
\end{minipage}
\caption{Quadratic model regression $1+\lambda(\mathcal{A})$ for low and high regularization $\beta$ with low and high model error $\gamma$.\label{quad_eig}}
\end{figure}

We simulate a quadratic system with $N=15$ and $M=20$. The entries of ${\bf A}$ and ground truth ${\bmath \theta}$ are generated from a standard normal distribution. In order to create model errors, we generate a sparse error matrix ${\bf A}_e$ that has $NM/3$ non-zero entries. These non-zero entries are also generated from a standard normal distribution. Afterwards ${\bf A}_e$ is scaled for a given $\gamma$ where $\gamma=\frac{\|{\bf A}_e\|}{\|{\bf A}\|}$. When we calculate ${\bf X}$ in (\ref{ex2_data}), we use ${\bf A}^\prime={\bf A}+{\bf A}_e$ instead of ${\bf A}$ and we add noise matrix ${\bf N}$ with signal to noise ratio (SNR) $\frac{\|{\bf X}\|}{\|{\bf N}\|}$ being equal to $30$.

For finding $\widehat{\bmath \theta}$ and for the evaluation of (\ref{ex2_sgrad}), (\ref{ex2_ftt}) and (\ref{ex2_ftx}), we use ${\bf A}$ and not the ground truth model ${\bf A}^\prime$. Moreover, we use ${\bmath \theta}=\widehat{\bmath \theta}$ for the evaluation of (\ref{ex2_sgrad}), (\ref{ex2_ftt}) and (\ref{ex2_ftx}).

In Fig. \ref{quad_eig}, we have shown $40$ eigenvalues $1+\lambda(\mathcal{A})$ (out of $D=300$). The remaining eigenvalues are all equal to $1$. We have used low and high regularization $\beta$ as well as low and high model error $\gamma$ to generate the eigenspectra in Fig. \ref{quad_eig}.  The key observation in Fig. \ref{quad_eig} is the behavior of the lowest $1+\lambda(\mathcal{A})$ at index $i=1$. With high model error and low regularization, this value is negative, clearly indicating that the estimation of ${\bmath \theta}$ has failed. Even with high regularization $\beta=100$, this value remains close to zero. This is due to the ambiguity of the possible solutions to ${\bmath \theta}$ and has nothing to do with the model ${\bf A}$ but more to do with the degrees of freedom \citep{cook1982residuals}. Increasing $\beta$ further would move all values of $1+\lambda(\mathcal{A})$ to $1$, as we have already seen in Fig. \ref{ridge_eig}, but that would not necessarily give the best solution to ${\bmath \theta}$.

In section \ref{sec:simul}, we will produce analogous plots of the eigenspectra of a radio interferometric calibration example. The inherent degeneracies, the model error and the regularization all act together in radio interferometric calibration and the examples we have presented in this section will be helpful to understand them in section \ref{sec:simul}.

\section{Radio interferometric simulations}\label{sec:simul}
We simulate an observation with $N=47$ stations, with the array geometry similar to the geometry presented in \cite{NCP2013}. Each calibration run uses $5$ time samples of integration time $10$ seconds each with each sample consisting of $N(N-1)/2=1081$ baselines. For consensus optimization, we use $P=24$ frequencies, spread over the range $[115,185]$ MHz. We use a Bernstein polynomial with $3$ terms as the consensus polynomial. The systematic errors are generated with the reference frequency at $f_0=150$ MHz, and smoothness in frequency enforced by polynomials similar to the one used by the simulations in \citep{DCAL}. To elaborate, we use a polynomial in frequency $f$, given as $\sum_{l=1}^{G} (\gamma_l+\jmath \delta_l) \left(\frac{f-f_0}{f_0}\right)^{(l-1)}$ where $\gamma_l,\delta_l$ are also drawn from a uniform distribution $\mathcal{U}(0,1)$ and $G=4$ (per each station and each direction in the sky). The systematic errors ${\bf { J}}^k_{pf}$ at $f=f_0$, for all directions in the sky $k$ and for all stations $p$, are generated to have complex Gaussian entries with zero mean and unit variance and are multiplied by the aforementioned polynomial in frequency to get values at $f$.

The sky model consists of $K$ point sources with fluxes in the range $[1.5,15]$ Jy and their positions are uniform randomly distributed over a field of view of 7$\times$7 square degrees. The sky also has an unknown (not part of the calibration model) Gaussian, at the phase center, with a spatial scale of a few arc minutes that affect the baselines that are shorter than about $2000$ wavelengths. The peak flux of the Gaussian is scaled as a fraction of the point source with lowest flux in the sky model, i.e., 1.5 Jy. The observed data consists of the $K$ point source signals (corrupted by systematic errors along each direction) plus the Gaussian, and finally, Gaussian noise with a signal to noise ratio of $50$ is also added to the simulated signal. The results presented in the following examples are at a frequency closest to the reference frequency, i.e., at $151.5$ MHz but we see similar results at other frequencies as well.

\begin{figure}
\begin{minipage}{0.98\linewidth}
\centerline{\epsfig{figure=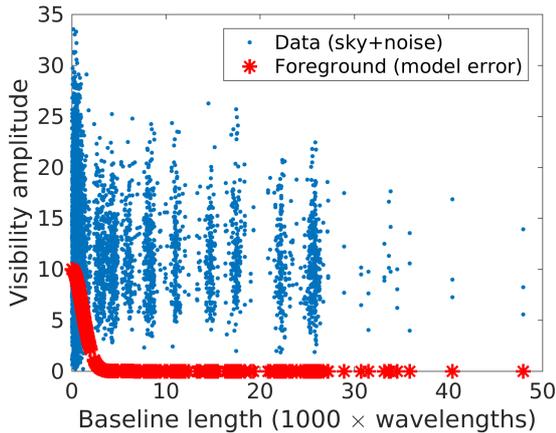,width=8.0cm}}
\end{minipage}
\caption{Simulated data and the model error which is a Gaussian. The model error only affects the short baselines.\label{calib_vis}}
\end{figure}

In Fig. \ref{calib_vis}, we show the simulated visibility's amplitude of the $XX$ correlation, plotted against the baseline length. We also show the contribution due to the model error (the Gaussian), which is hidden in the simulated data and which is affecting only the short baselines. 

We perform calibration using $5$ time samples for various values of $K$, $\rho$ and the peak value of the Gaussian foreground. Unless stated otherwise, for each situation where $K$, $\rho$ or the Gaussian foreground amplitude is varied, we perform $10$ Monte Carlo runs, where we randomly generate the source positions in the sky, the systematic errors and the additive noise. What is shown in the following is the averaged eigenspectra over all $10$ Monte Carlo runs. When $\rho=0$, consensus optimization is not performed, hence it corresponds to traditional calibration. Based on our previous work \citep{EUSIPCO2016}, we use $\rho=500$ for a source of peak flux $15$ Jy and the values of $\rho$ for other sources are scaled linearly according to their flux. During traditional calibration we use $100$ iterations and during distributed calibration, we use $40$ ADMM iterations and $20$ inner iterations for (\ref{step1}). 

\begin{figure}
\begin{minipage}{0.98\linewidth}
\centerline{\epsfig{figure=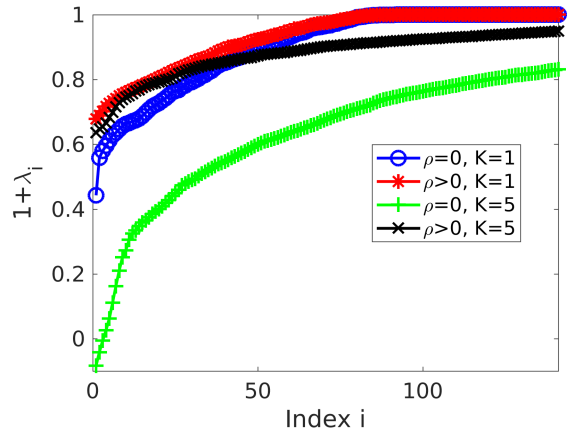,width=8.0cm}}
\end{minipage}
\caption{Eigenvalues $1+\lambda(\widetilde{\mathcal{A}})$ for traditional calibration and consensus optimization with direction independent ($K=1$) and direction dependent calibration ($K=5$).\label{calib_eigK}}
\end{figure}

\begin{figure}
\begin{minipage}{0.98\linewidth}
\centerline{\epsfig{figure=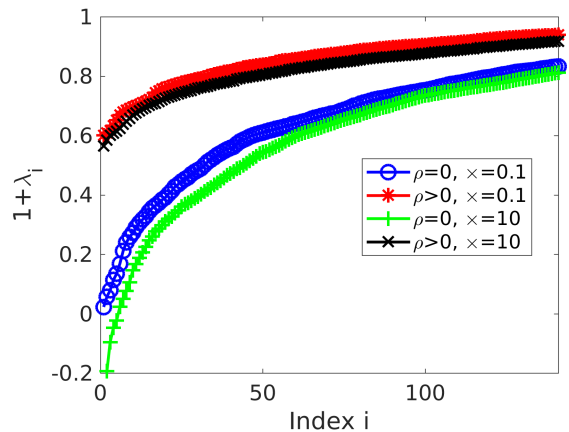,width=8.0cm}}
\end{minipage}
\caption{Eigenvalues $1+\lambda(\widetilde{\mathcal{A}})$ for calibration along $K=5$ directions with various values of model error. The model error spatial scale is kept constant while the amplitude is scaled as a fraction of the flux of the weakest source, i.e., $1.5$ Jy.\label{calib_eigG}}
\end{figure}

With $5$ time samples and $1081$ baselines, we have $D=5\times 1081\times 8$ and finding eigenvalues of the full $\mathcal{A}$ in (\ref{deteigs}) is expensive. We select a subset of rows and columns from the Jacobian, and we only consider the mapping of the real part of $XX$ correlation (from input to the output residual), averaged over the $5$ time samples. Therefore, we have a reduced matrix $\widetilde{\mathcal{A}}$ of size $1081\times 1081$. We show the smallest $3N$ values of $1+\lambda(\widetilde{\mathcal{A}})$ in Figs. \ref{calib_eigK} and \ref{calib_eigG} for various calibration scenarios. 

In Fig. \ref{calib_eigK}, we have kept the peak of the foreground Gaussian flux at $\times 1$, i.e., at 1.5 Jy and we have varied $K$. We show results for calibration along one direction ($K=1$) and for calibration along $K=5$ directions. In both cases, we show the performance of calibration without consensus optimization ($\rho=0$) and with consensus optimization ($\rho>0$). One might conclude that calibration over multiple directions ($K=5$) gives worse performance than calibration over a single direction ($K=1$), especially for $\rho=0$. This result needs further explanation. What is shown in Fig. \ref{calib_eigK} in fact is a measure of the distortion of the residual statistics due to calibration, with eigenspectra that are much lower than $1$ indicating more distortion. This agrees with what has been observed, for instance in terms of the suppression of diffuse foregrounds \citep{Patil2016} where the suppression increases with increasing $K$. However, this is not the only criterion for selecting $K$ for calibration. Depending on an observation, we might have no other choice but to subtract all interfering sources in the sky to reach the desired result. Nonetheless, we should keep in mind that this comes at a price, i.e., the statistical distortion to the residual. This distortion can manifest in many ways including signal suppression, noise increase and over fitting.

In Fig. \ref{calib_eigG}, we consider calibration along $K=5$ directions. We have changed the peak value of the Gaussian foreground (that acts as a model error) as a fraction ($\times 0.1$ and $\times 10$) of the weakest source in the sky model, i.e., $1.5$ Jy. It is clear that conventional calibration ($\rho=0$) is affected more by errors in the sky model. The eigenspectra for distributed calibration ($\rho>0$) also get affected, but to a lesser extent. We can compare the well known regression problems presented in Figs. \ref{ridge_eig} and \ref{quad_eig} to the results based on radio interferometric calibration in Figs. \ref{calib_eigK} and \ref{calib_eigG}. To summarize, we can consider radio interferometric calibration as a non-linear regression as well and try to improve our understanding of its performance by using the knowledge of non-linear regression \citep{cook1982residuals,Hempel86}.

From Figs. \ref{calib_eigK} and \ref{calib_eigG} we can conclude that just by looking at the lowest values of $1+\lambda(\widetilde{\mathcal{A}})$, we can have a reasonable measure on the statistical performance of calibration. This is especially relevant when manual checking is not possible, for instance in pipeline processing. Finding only a few eigenvalues also cuts down the computational cost because we can use iterative algorithms \citep{ARPACK}. On the other hand, we also observe that when $\rho>0$, the relative change in the values of $1+\lambda(\widetilde{\mathcal{A}})$  is small. To observe subtle differences in performance, we can produce {\em influence maps} as discussed in section \ref{sec:model}. 

\begin{figure}
\begin{minipage}{0.98\linewidth}
\begin{minipage}{0.48\linewidth}
\centerline{\epsfig{figure=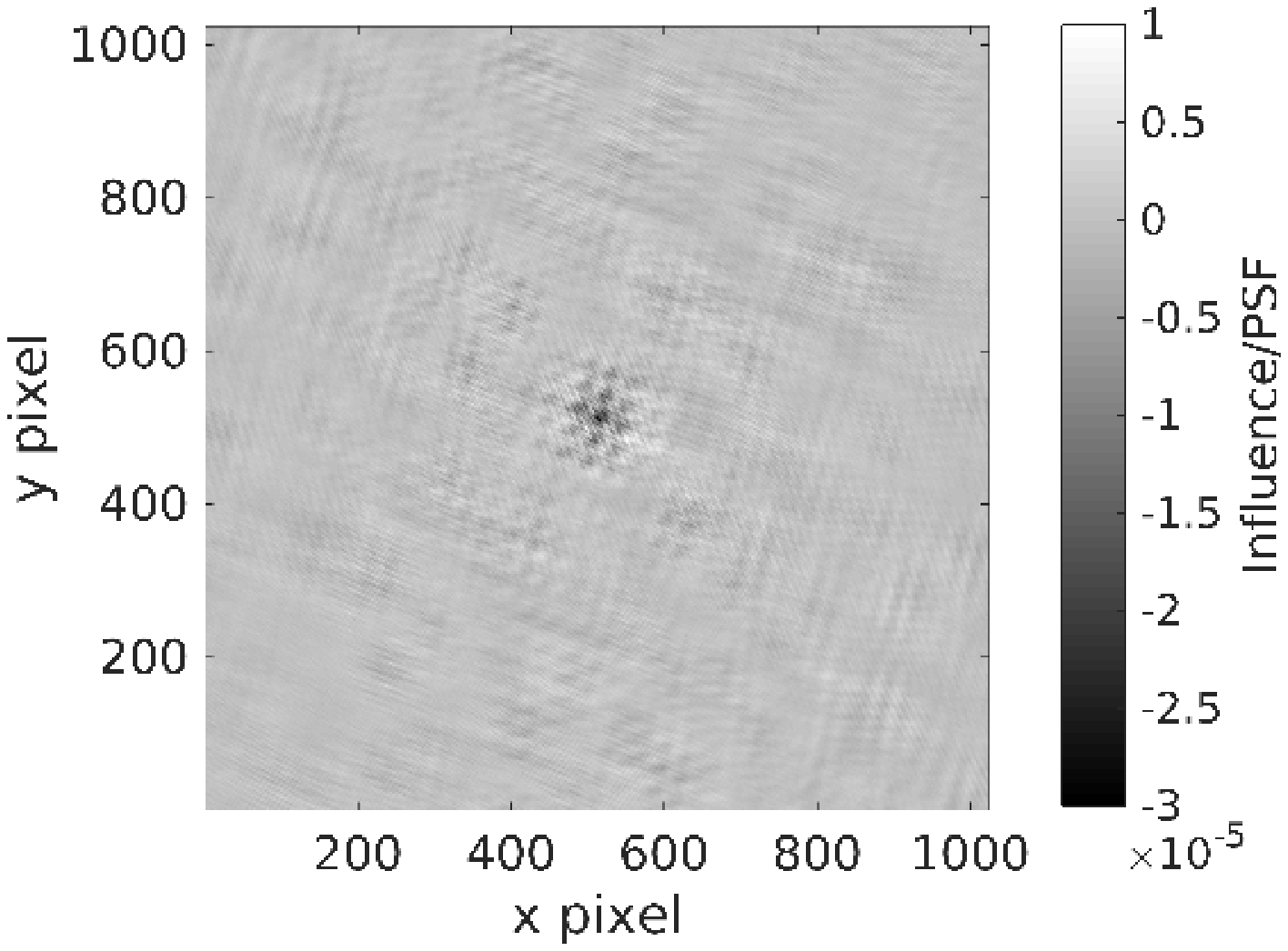,width=4.0cm}}
\end{minipage}
\begin{minipage}{0.48\linewidth}
\centerline{\epsfig{figure=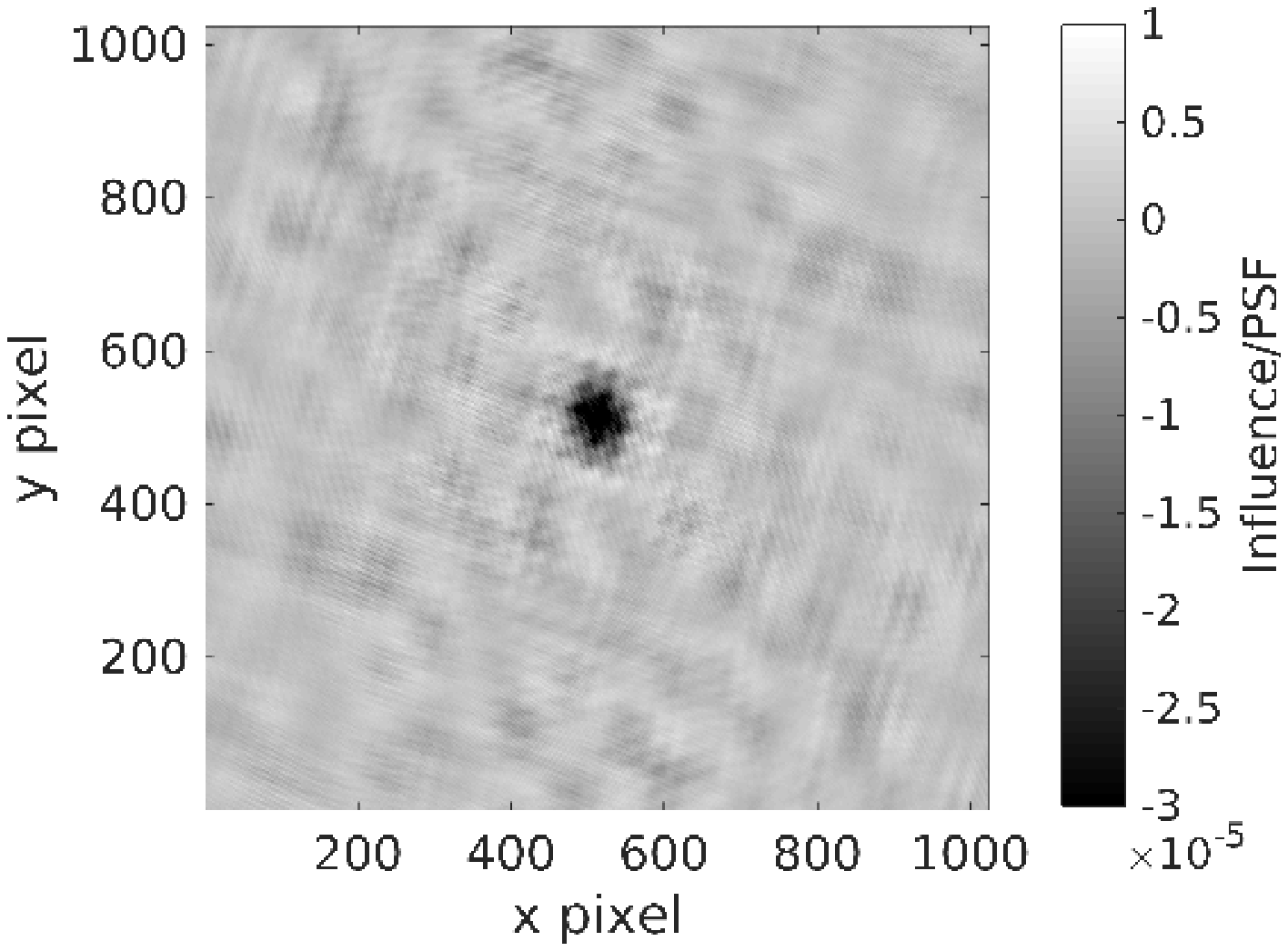,width=4.0cm}}
\end{minipage}
\end{minipage}
\caption{Influence maps (real $XX$) for calibration along $K=5$ directions using a short integration of about $21$ minutes. Both images have same intensity scale (left) low model error, high noise (right) high model error, low noise.\label{influence_map}}
\end{figure}

In Fig. \ref{influence_map}, we show the influence map made by averaging elements in (\ref{Jmapping}). Instead of the writing back the residual ${\bf y}$ as output (in which case each $y_i$ is written as the $i$-th data point), we find the average of $\frac{\partial y_i}{\partial x_j}$ (averaged over all values of $j$) and write this as the $i$-th data point. To have enough sampling in the Fourier space, we perform calibration on a dataset with $125$ time samples (which is about 21 minutes of integration) using $5$ time samples per each calibration. The sky model ($K=5$) and the foreground error Gaussian is kept fixed throughout the total integration but the systematic errors and the noise are re-generated for each $5$ time samples.

We show the dirty images with a pixel size of $2^{\prime\prime} \times 2^{\prime\prime}$ in Fig. \ref{influence_map} for two different scenarios. On the left, we have high noise (with SNR$=5$) while model error is low ($\times=0.1$) and on the right, we have lower noise (with SNR=$50$) and model error is higher ($\times=1$). We see that averaging $\frac{\partial y_i}{\partial x_j}$ enhances the systematic effect due to the model error while the systematic effects due to the noise are more random and thus are less prominent. Further investigations using influence mapping is left as future work.

\section{Conclusions}\label{sec:conc}
We have developed an analytic framework to measure the statistical performance of direction dependent, distributed radio interferometric calibration. We have also applied our analysis to two well known regression problems. By comparing the eigenspectra of the Jacobian of the example regression problems with  the one obtained in radio interferometric calibration, we see that essentially all are similar. Therefore, there is opportunity to re-use ideas developed in statistical regression to study and improve calibration. As a simple diagnostic or a measure of confidence on the performance of calibration, we can use (a subset of) the eigenspectrum of the Jacobian and going further, influence mapping can visualize more subtle effects. The techniques presented in this paper can be extended in many ways. For instance, since we have determined the relationship between the input and output PDFs, we can recover the power spectra of the original signal (input) by correcting the residual power spectra. Furthermore, we can apply the same techniques to machine learning problems in other disciplines such as deep learning \citep{Sokolic2017} to detect model errors or over fitting.

\section*{Acknowledgments}
This work is supported by Netherlands eScience Center (project DIRAC, grant 27016G05). We thank the anonymous reviewers for valuable comments.

\begin{appendix}
\section{Derivative of corrected data}\label{app:Corr}
We use $d ({\bf Z}^{-1})=-{\bf Z}^{-1}(d{\bf Z}) {\bf Z}^{-1}$ \citep{Mdiff} to get
\beq
d({\bf A}_p {\bf J}_f)^{-1} =-({\bf A}_p {\bf J}_f)^{-1} {\bf A}_p (d{\bf J}_f) ({\bf A}_p {\bf J}_f)^{-1}
\eeq
and 
\beq
d({\bf A}_q {\bf J}_f)^{-H} =-\left(({\bf A}_q {\bf J}_f)^{-1} {\bf A}_q (d{\bf J}_f) ({\bf A}_q {\bf J}_f)^{-1}\right)^{H}.
\eeq
Using the above, we get the differential of (\ref{corr}) as
\beq
d{\bf Q}= -({\bf A}_p {\bf J}_f)^{-1} {\bf A}_p (d{\bf J}_f) {\bf Q}- {\bf Q}(d{\bf J}_f)^H{\bf A}_q^T({\bf A}_p {\bf J}_f)^{-H}.
\eeq
Vectorizing both sides and using definition 4 of \cite{Mdiff} we get the derivative as (\ref{Qderiv}).

\section{Selecting subsets of data}\label{app:Subset}
We can rewrite (\ref{Jmapping}) in block matrix form as
\beq
\mathcal{J} =\left[
\begin{array}{cc}
{\bf A} & {\bf B}\\
{\bf C} & {\bf D}
\end{array}
\right]
\eeq
where ${\bf A}$ is a square matrix that includes the rows and columns that we select as a subset of interest. Note that ${\bf D}$ is also a square matrix but ${\bf B}$ and ${\bf C}$ can have any conforming shape. From (\ref{Reffderiv}) we also see that $\mathrm{vec}\left( \frac{\partial{\bf V}_{pqf}} {\partial x_{p^\prime q^\prime r}} \right)$ is only contributing a $1$ to the diagonal entries of $\mathcal{J}$ and the off diagonal entries are much smaller than $1$ at the converged solution. Therefore, we generally have 
\beq \label{appnorm}
\|{\bf A}\| \gg \| {\bf B} \|, \|{\bf A}\| \gg \| {\bf C} \|,
 \|{\bf D}\| \gg \| {\bf B} \|, \|{\bf D}\| \gg \| {\bf C} \|.
\eeq
Another way to say (\ref{appnorm}) is that the influence of $x_i$ on $y_i$ is much larger than the influence of $x_i$ on $y_j$ when $i\ne j$.
We can rewrite the determinant of $\mathcal{J}$ in block matrix form as 
\beq
|\mathcal{J}| =| {\bf A} - {\bf B} {\bf D}^{-1} {\bf C} | |{\bf D}| 
\eeq
and using (\ref{appnorm}) we have $|\mathcal{J}| \approx |{\bf A}||{\bf D}|$ or in other words, $|\mathcal{J}| \propto |{\bf A}|$ for the subset of selected rows and columns. Therefore, we get a similar performance by studying $|{\bf A}|$ as we get by studying $|\mathcal{J}|$.
\end{appendix}
\bibliographystyle{mnras}
\bibliography{references}
\bsp
\label{lastpage}
\end{document}